\documentclass[sigconf]{acmart}
\usepackage{amsmath,amsfonts}
\usepackage{algorithmic}
\usepackage{graphicx}
\usepackage{textcomp}
\usepackage{hyperref}
\usepackage{pifont}
\usepackage{caption}
\usepackage{geometry}
\usepackage{longtable}
\usepackage{array}
\usepackage{booktabs}
\usepackage{enumitem}
\usepackage{multirow}
\usepackage{listings}
\usepackage{todonotes}
\usepackage{tcolorbox}
\usepackage{lipsum}
\usepackage{algorithm}
\usepackage{colortbl}
\usepackage{rotating}
\usepackage{booktabs}
\usepackage{multicol}
\usepackage{titlesec}
\usepackage{xcolor}
\usepackage[utf8]{inputenc}
\usepackage{lmodern}
\usepackage[T1]{fontenc}

\usepackage{placeins}

\usepackage{colortbl}
\usepackage{pifont}
\usepackage{graphicx}

\definecolor{bluebox}{HTML}{E3F2FD}   
\definecolor{redbox}{HTML}{FFEBEE}    
\definecolor{greenbox}{HTML}{E8F5E9}  
\definecolor{graybox}{HTML}{ECEFF1}   
\definecolor{titlecolor}{HTML}{1A237E} 
\definecolor{codebg}{HTML}{FAFAFA}    

\lstdefinestyle{customc}{
  belowcaptionskip=1\baselineskip,
  breaklines=true,
  frame=L,
  xleftmargin=\parindent,
  language=C,
  showstringspaces=false,
  numbers=left,               
  numberstyle=\tiny\color{gray}, 
  stepnumber=1,
  basicstyle=\tiny\ttfamily, 
  keywordstyle=\bfseries\color{green!40!black},
  commentstyle=\itshape\color{purple!40!black},
  numbersep=8pt,
  identifierstyle=\color{blue},
  stringstyle=\color{orange},
}

\newboolean{showcomments}
\setboolean{showcomments}{true}
\ifthenelse{\boolean{showcomments}}
{
}

\definecolor{lightgreen}{HTML}{98FB98}
\definecolor{lightyellow}{HTML}{FFFF99}
\definecolor{lightred}{HTML}{FFB6C1}

\newlist{numpar}{enumerate}{1}
\setlist[numpar,1]{label=\arabic*)}

\renewcommand\footnotetextcopyrightpermission[1]{} 
\acmConference[]{}{}{} 

\definecolor{bgcolor}{rgb}{0.95,0.95,0.95}  
\definecolor{codeblue}{rgb}{0.0, 0.0, 0.6}  
\definecolor{codered}{rgb}{0.6, 0.0, 0.0}   
\definecolor{codegray}{rgb}{0.5, 0.5, 0.5}  

\usepackage{graphicx}
\usepackage{listings}
\usepackage{xcolor}
\usepackage{caption}
\usepackage{framed} 

\definecolor{codegreen}{RGB}{0,128,0}      
\definecolor{codered}{RGB}{204,0,0}        
\definecolor{codegray}{RGB}{128,128,128}   
\definecolor{codeblue}{RGB}{0,102,153}     
\definecolor{stringcolor}{RGB}{208,76,0}   
\definecolor{bgcolor}{RGB}{248,248,248}    
\definecolor{linecolor}{RGB}{220,220,220}  
\definecolor{addbg}{RGB}{230,255,230}      
\definecolor{delbg}{RGB}{255,230,230}      

\lstdefinestyle{cstylediff}{
  language=C++,
  backgroundcolor=\color{bgcolor},
  basicstyle=\ttfamily\scriptsize,
  keywordstyle=\color{codeblue}\bfseries,
  commentstyle=\color{codegray}\itshape,
  stringstyle=\color{stringcolor},
  numbers=left,
  numberstyle=\tiny\color{codegray},
  stepnumber=1,
  numbersep=6pt,
  frame=single,
  rulecolor=\color{linecolor},
  tabsize=2,
  breaklines=true,
  breakatwhitespace=false,
  showspaces=false,
  showstringspaces=false,
  captionpos=b,
  aboveskip=4pt,
  belowskip=4pt,
  xleftmargin=2em,
  framexleftmargin=1.5em,
  keepspaces=true,
  upquote=true,
  lineskip=-0.1pt,          
  basewidth={0.5em, 0.4em}, 
  columns=fixed,            
}

\begin{document}

\title{Reasoning with LLMs for Zero-Shot Vulnerability Detection}

\author{Arastoo Zibaeirad}
\affiliation{%
  \institution{University of North Carolina at Charlotte}
  \city{Charlotte}
  \state{NC}
  \country{USA}
}
\email{azibaeir@charlotte.edu}

\author{Marco Vieira}
\affiliation{%
  \institution{University of North Carolina at Charlotte}
  \city{Charlotte}
  \state{NC}
  \country{USA}
}
\email{marco.vieira@charlotte.edu}

\begin{abstract}
Automating software vulnerability detection (SVD) remains a critical challenge in an era of increasingly complex and interdependent software systems. Despite significant advances in Large Language Models (LLMs) for code analysis, prevailing evaluation methodologies often lack the \textbf{context-aware robustness} necessary to capture real-world intricacies and cross-component interactions. To address these limitations, we present \textbf{VulnSage}, a comprehensive evaluation framework and a dataset curated from diverse, large-scale open-source system software projects developed in C/C++. Unlike prior datasets, it leverages a heuristic noise pre-filtering approach combined with LLM-based reasoning to ensure a representative and minimally noisy spectrum of vulnerabilities. The framework supports multi-granular analysis across function, file, and inter-function levels and employs four diverse zero-shot prompt strategies: Baseline, Chain-of-Thought, Think, and Think \& Verify.
Through this evaluation, we uncover that structured reasoning prompts substantially improve LLM performance, with Think \& Verify reducing ambiguous responses from 20.3\% to 9.1\% while increasing accuracy. We further demonstrate that code-specialized models consistently outperform general-purpose alternatives, with performance varying significantly across vulnerability types, revealing that no single approach universally excels across all security contexts. \textbf{Link to dataset and codes:} \href{https://github.com/Erroristotle/VulnSage.git}{https://github.com/Erroristotle/VulnSage.git}

\end{abstract}


\maketitle

\section{Introduction}

Automated software vulnerability detection (SVD) refers to using software tools and techniques to automatically identify weaknesses or flaws in code that could be exploited by attackers. 
Such tools help developers and security professionals to proactively address vulnerabilities and improve security.

Program analysis techniques have been the gold standard for SVD. These techniques utilize static and dynamic approaches, but are often challenged by false positive rates, complex dependency graphs, and limited coverage \cite{manes2019art, klees2018evaluating, pereira2021machine, li2024llm}. Such limitations are intensified in modern software environments, where vulnerabilities frequently arise from complex interactions between components and subtle logical errors.

Significant advancements have been made in Large Language Models (LLMs) \cite{chen2021evaluating, dubey2024llama, guo2025deepseek}, which demonstrate promising capabilities in code understanding and semantic reasoning \cite{guo2024deepseek, roziere2023code}. Although LLMs offer superior semantic reasoning capabilities compared to program analysis tools, they face notable limitations in generalizing to real-world vulnerabilities, maintaining robustness against noisy and incomplete data, and reliably detecting both common (e.g., CWE-119, CWE-20) and complex vulnerabilities. 

Assembling vulnerability datasets has traditionally relied on synthetic code or on crawling repositories such as the National Vulnerability Database (NVD) for fix commits \cite{bhandari2021cvefixes, chakraborty2021deep, pereira2022software, chen2023diversevul}. These works typically extract modified files or functions and employ pre-commit analysis to isolate security-related changes. However, this method often introduces significant noise from unrelated modifications (e.g., refactoring, documentation updates) \cite{iannone2022secret, croft2022noisy}. Prior efforts to mitigate noise have relied either on manual annotation, which is labor-intensive and non-scalable, or on simple heuristic filters that do not provide a quantitative assessment of noise \cite{zhou2019devign, hommersom2024automated, gao2023far, siddiq2022securityeval, fan2020ac, thomas2022dynamic}.

To address these challenges, we propose \textit{VulnSage}, an evaluation framework and dataset designed for rigorous benchmarking LLMs in SVD. VulnSage consists of 593 vulnerabilities across 52 CWE categories, curated from large-scale open-source system software projects, including the Linux Kernel, Mozilla, Gecko-dev, and Xen. The dataset is constructed using a structured pipeline that, by crawling CVEDetails,  retrieves vulnerability metadata (CVE, CWE, commit hashes), extracts relevant code changes, and organizes them into labeled vulnerability-patch code block pairs. To ensure dataset reliability, VulnSage employs a two-stage noise mitigation strategy:
(1) heuristic pre-filtering eliminates noisy commits, such as those involving excessive file modifications, refactoring changes, missing ground-truth CWE labels, or non-security-related updates; and (2) an LLM-based reasoner assigns a quantitative noise score (0 - 100\%) to the remaining samples by analyzing Git diffs and commit descriptions, ensuring a cleaner and more representative dataset. Our dataset is publicly available and is designed to be extensible, allowing the integration of additional projects, CWEs, and vulnerabilities.

Our evaluation workflow employs a multi-granular approach, analyzing vulnerabilities beyond isolated function- and file-level perspectives to provide a more comprehensive assessment of LLMs' detection capabilities. Instead of treating functions or files as independent units, we construct code blocks that incorporate paths of modified files and the inter-function relationships within each file, making the dataset more representative of modern software structures. This approach reflects real-world vulnerability patterns, where security flaws often arise from interactions between multiple functions within a file or across files rather than existing in isolation. By capturing these dependencies, execution flows, and shared resources, we ensure that LLMs are evaluated on their ability to detect vulnerabilities not only within individual functions or files but also in their contextual interplay. To further ensure a rigorous evaluation, we assess potential training data leakage by testing LLMs on a curated subset of vulnerabilities discovered after their training cutoff dates. This allows us to distinguish genuine zero-shot reasoning capabilities from potential memorization of previously seen security flaws \cite{carlini2022quantifying, zibaeirad2024comprehensive}.

We employ four sophisticated zero-shot prompting strategies to evaluate the intrinsic vulnerability detection capabilities of LLMs without relying on task-specific fine-tuning or labeled data. These include a Baseline strategy, which uses simple binary classification; the Chain-of-Thought (CoT) strategy, which incorporates structured, step-by-step reasoning; the Think strategy, which requires explicit logical analysis within a structured framework; and the Think \& Verify strategy, which adds self-verification, confidence scoring, and severity assessment. This diverse set of prompts enables us to rigorously analyze how effectively twelve distinct LLMs identify vulnerabilities across various contexts and systematically examine potential biases toward specific CWE classes. This comprehensive framework shows the models' capabilities in reasoning about software security while exposing their limitations when confronted with complex, cross-component vulnerabilities. 

Our evaluation reveals key insights into LLM performance in vulnerability detection. Despite dataset noise, structured reasoning strategies such as Think \& Verify improve robustness, mitigating the impact of irrelevant modifications and reducing ambiguous responses from 20.3\% to 9.1\%. Our data leakage analysis confirms that models rely on reasoning rather than memorization, with Think \& Verify achieving 67.5\% accuracy even on vulnerabilities discovered after training cutoffs. Additionally, we find that detection accuracy varies with code granularity—while multi-function and multi-file contexts improve vulnerability detection, patch verification becomes more challenging due to increased complexity. 

In short, our key contributions are as follows: 
\begin{itemize}[leftmargin=*, labelsep=5mm]
\item Introduce VulnSage, a novel evaluation framework accompanied by a curated dataset drawn from diverse, large-scale open-source projects developed in C/C++. The dataset includes 593 vulnerabilities across 52 CWE categories.
\item Multi-level code analysis that captures vulnerabilities emerging from component interactions by constructing code blocks with file paths and inter-function relationships, better reflecting real-world security scenarios.

\item Four zero-shot prompting strategies (Baseline, CoT, Think, Think \& Verify) designed to evaluate structured reasoning, evidence-based analysis, and self-verification in vulnerability detection tasks.

\item Systematic evaluation revealing LLM limitations across different CWE classes, code granularities, and reasoning approaches, highlighting performance patterns in zero-shot vulnerability detection.
\end{itemize}

The rest of the paper is organized as follows. \textbf{Section \ref{section:related_work}} reviews related work on vulnerability detection and LLM evaluation. \textbf{Section \ref{section: Experimental Setup}} introduces the \textit{VulnSage} framework, covering dataset collection, noise mitigation, and evaluation methodology. \textbf{Section \ref{section:result}} presents experimental results, analyzing LLM performance, prompt effectiveness, data leakage, noise impact, and CWE biases. \textbf{Section \ref{section:threats}} discusses threats to validity, and \textbf{Section \ref{section:conclusion}} concludes with key findings and future directions.

\section{Related Work}
\label{section:related_work}

\subsection{Vulnerability Detection Datasets}

Existing datasets often concentrate on a single granularity, be it at the line level \cite{fu2022linevul}, function level \cite{zhou2019devign, chakraborty2021deep, chen2023diversevul}, file level \cite{liu2024vuldetectbench}, or repository level \cite{zhou2024comparison}, thus consequently failing to represent real-world vulnerabilities that arise from cross-function dependencies and multi-file interactions.
To improve dataset reliability, previous works have employed manual annotation or simple heuristic filters to remove noise \cite{zhou2019devign, hommersom2024automated, gao2023far}, but these methods lack scalability and quantitative noise assessment. 

VulnSage addresses such limitations by introducing a multilevel dataset curation strategy that integrates: (1) multigranular code representations incorporating function and file interactions, (2) automated two-stage noise mitigation, combining heuristic filtering with LLM-based noise quantification, and (3) an evaluation of LLMs on vulnerabilities discovered after their training cutoff dates to analyze potential data leakage effects. These improvements allow a more realistic and reliable evaluation of LLM-driven vulnerability detection.

\subsection{Evaluation of LLMs in SVD}
The advent of LLMs has significantly advanced SVD by enhancing code comprehension and semantic reasoning \cite{ullah2024llms, zibaeirad2024vulnllmeval, abshari2024llm, abshari2025survey,babaey2025gensqli, haghighi2024eye, babaey2025genxss,alam2024assess, babaey2025detecting}. However, most existing evaluations rely on synthetic datasets or isolated function-level analyses \cite{khare2023understanding}, failing to capture the complexities of cross-component interactions and the impact of noise on detection performance. Additionally, many benchmarks suffer from data leakage, as test sets often contain code snippets present in LLM training data \cite{sallou2024breaking, wu2023effective}, raising concerns about models memorizing vulnerabilities rather than reasoning about them.
\begin{figure*}[htbp]
    \centering
    \includegraphics[width=1.0\linewidth]{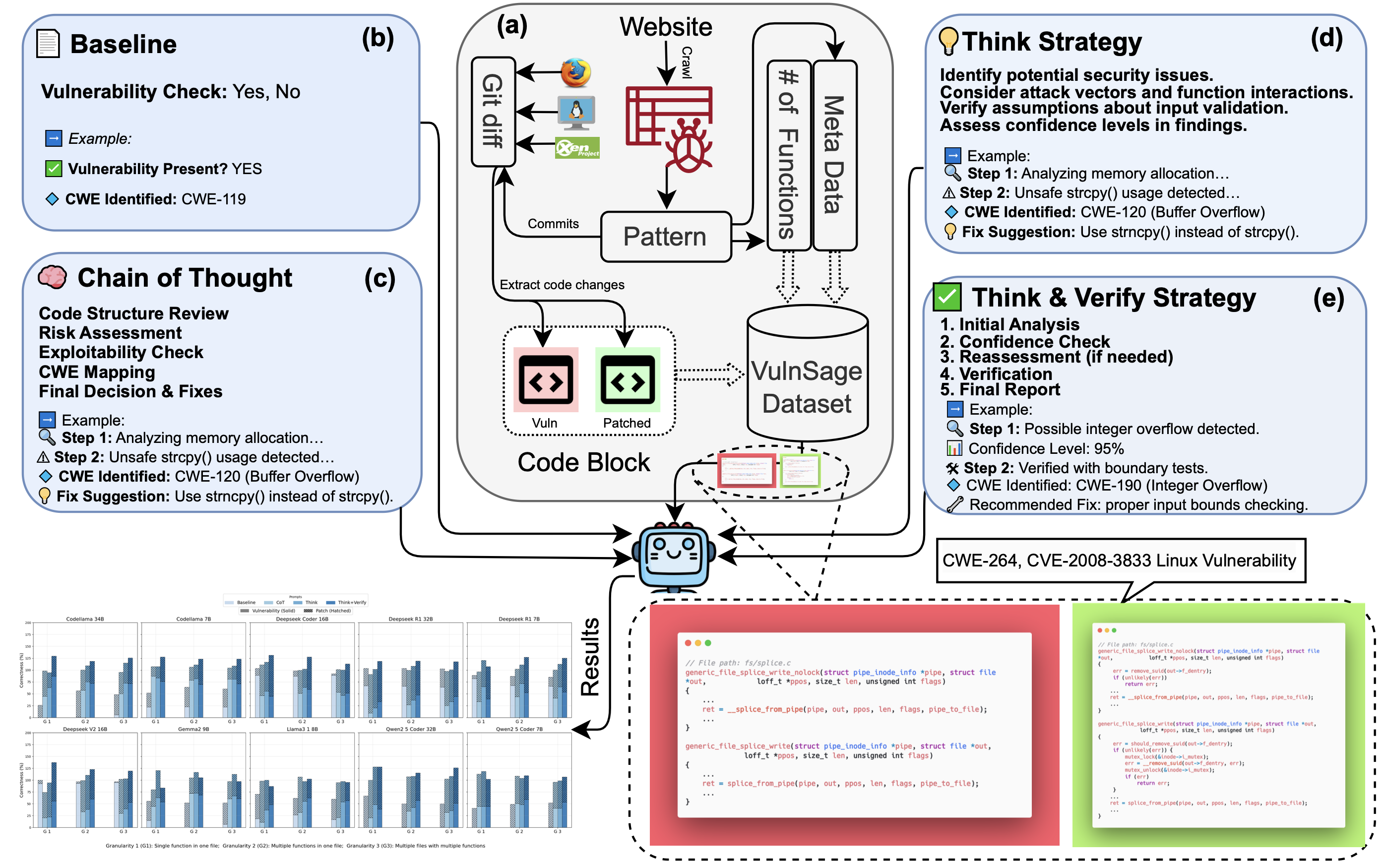}
    \caption{VulnSage Architecture}
    \label{fig:vulnsage}
\end{figure*}
Beyond data leakage, prior evaluations primarily focus on common vulnerability patterns while neglecting a diverse range of CWE types and real-world CVEs, limiting their applicability to security-critical systems \cite{ullah2024llms}. Furthermore, existing studies do not systematically assess LLMs using structured reasoning prompts, such as Think and Think \& Verify, to evaluate how explicit logical reasoning affects performance in both vulnerability detection and patch verification. This omission leaves a critical gap in understanding how LLMs analyze security flaws beyond binary classification.

VulnSage addresses these shortcomings by evaluating LLMs across multiple granularity levels, providing a more realistic assessment of their ability to detect subtle, system-level vulnerabilities. Additionally, to assess whether models rely on memorization, we curated a subset of test data consisting of vulnerabilities discovered after the LLMs' training cutoff dates. With a dataset comprising 593 vulnerabilities spanning 52 CWEs and 491 CVEs, VulnSage delivers a comprehensive, diverse, and rigorous evaluation framework that better reflects real-world security challenges in C/C++, while integrating structured reasoning prompts to uncover the reasoning capabilities of pre-trained LLMs in security-critical contexts.

\section{VulnSage Framework}
\label{section: Experimental Setup}
This section introduces the VulnSage Framework, our comprehensive approach for evaluating LLMs on real-world software vulnerabilities. As shown in Figure \ref{fig:vulnsage}, we specifically focus on authentic vulnerabilities that exhibit complex interactions across files and functions, making their detection particularly challenging for automated tools. To enable this evaluation, we carefully curated a dataset of real-world vulnerabilities with their complete context. Our assessment methodology employs four distinct prompting strategies, with three specifically designed to test the reasoning capabilities of LLMs when analyzing vulnerability patterns.

\subsection{Dataset Collection}
\label{subsection:dataset}
To build VulnSage, we collected vulnerability metadata from the Linux Kernel, Mozilla, and Xen, as they are widely used, security-critical system software with diverse code structures and real-world vulnerabilities. As shown in Figure~ \ref{fig:vulnsage}, we used CVEDetails \cite{cvedetails} to gather commit hashes, CVEs, and CWEs, enabling us to extract vulnerable code blocks and their corresponding patched versions.

Once the metadata were collected, we extracted vulnerable and patched code blocks from the project's repository by analyzing the diff files of each commit. This allows us to identify added and deleted lines, as well as the files and functions that have been modified. If a change occurs within a function, we extract both the vulnerable and patched versions of the entire function to preserve contextual integrity. If the change is outside a function, such as a global variable modification, we include it at the top of the respective vulnerable or patched code block to maintain correctness. Since we apply specific filtering criteria to ensure data quality, we collect a substantial but curated subset of vulnerabilities from each project.

To ensure the quality of the dataset, we implement a noise filtration process. We automatically remove samples with excessive modifications, defined as changes spanning more than 500 lines of code, to minimize noise and prevent exceeding LLM context length limits. Samples lacking a ground-truth CWE are also excluded to maintain proper classification. Additionally, we filter out cases where a patch introduces a completely new function absent in the vulnerable code or removes an entire function, resulting in an empty vulnerable or patched code block. These filtering criteria ensure a high-quality dataset that accurately represents real-world vulnerabilities while remaining suitable for LLM evaluation.

Since eliminating all noise from security patches is impractical, we implemented a systematic approach to quantify residual noise using Deepseek R1 7B. For each commit in our dataset, we provided the model with the complete Git diff and commit description, then instructed it to estimate the proportion of non-security-related modifications on a 0–100\% scale (0\% representing purely security-relevant changes, 100\% indicating mostly unrelated modifications). The model followed a structured process: analyzing the commit's stated intent, reviewing each code modification, and assigning a quantified noise score with supporting reasoning. To validate this methodology, we manually reviewed a random 2\% sample of the dataset, finding strong agreement between LLM-assigned scores and human assessments across all examined commits. This validation confirms that our noise quantification approach effectively discriminates between security-critical fixes and peripheral changes, providing valuable context for interpreting model performance. We further analyze how varying levels of commit noise affect LLM accuracy in Section \ref{sec:impactofnoise}.

To capture cross-function vulnerabilities, we ensure that interactions between functions within and across files are retained. When extracting modified code, we collect all functions affected in a commit, preserving their relationships and execution flow. If a change spans multiple functions or files, all relevant code is included in the extracted vulnerable and patched code blocks. This structured representation allows LLMs to analyze security flaws in context rather than treating functions as isolated entities.

In constructing VulnSage, we prioritize diversity and representativeness, encompassing 52 CWEs and 593 total vulnerabilities. As detailed in Table~\ref{tab:dataset_characterization}, the dataset spans a wide range of vulnerabilities, from minor modifications involving a few lines to extensive patches affecting hundreds of lines of code. This variation in scale and complexity ensures that VulnSage serves as a rigorous and comprehensive benchmark for evaluating LLMs in SVD tasks, providing realistic challenges that align with real-world security concerns.
\begin{table}[htbp]
    \centering
    \caption{Dataset Characterization}
    \renewcommand{\arraystretch}{0.8}
    \resizebox{\columnwidth}{!}{
    \begin{tabular}{l|r||l|r}
        \hline
        \multicolumn{2}{c||}{\textbf{Vulnerability Characteristics}} & \multicolumn{2}{c}{\textbf{Code Structure}} \\
        \hline
        Total Vulnerabilities & 593 & Avg. Files Changed & 2.71 \\
        Unique CVEs & 491 & Median Files Changed & 1.00 \\
        Unique CWEs & 52 & Avg. Functions Changed & 18.42 \\
        Most Common CWE & CWE-119 & Median Functions Changed & 5.00 \\
        Year Range & 2002--2019 & Avg. Lines in Vulnerable Code & 296.79 \\
        & & Avg. Lines in Patched Code & 304.92 \\
        \hline
        \multicolumn{2}{c||}{\textbf{Patch Metrics}} & \multicolumn{2}{c}{\textbf{Granularity Distribution}} \\
        \hline
        Avg. Lines Added & 35.39 & Single function (G1) & 27 \\
        Avg. Lines Deleted & 20.12 & Multiple functions, single file (G2) & 244 \\
        Median Lines Added & 8.00 & Multiple files \& functions (G3) & 322 \\
        Median Lines Deleted & 5.00 \\
        \hline
    \end{tabular}
    }
    \label{tab:dataset_characterization}
\end{table}
\subsection{Prompt Templates}
\label{subsection:prompts}
To systematically evaluate LLMs for SVD, we employ four distinct prompt strategies, each targeting different aspects of reasoning and detection accuracy. These prompts range from simple binary classification to structured multi-step reasoning and self-verification mechanisms, allowing for a nuanced analysis of LLMs’ strengths and limitations in vulnerability detection.

The \textbf{Baseline Prompt} is a straightforward, direct approach that asks the model whether a specific CWE exists in the given code using a binary (YES/NO) classification without requiring explanations or reasoning. This serves as a control to evaluate how well LLMs can detect vulnerabilities without additional reasoning scaffolding.

The \textbf{CoT Prompt} \cite{wei2022chain, zhang2022automatic} builds upon structured reasoning principles by guiding the model through a six-step analytical process: (1) code structure analysis, (2) vulnerability pattern recognition, (3) attack surface and risk assessment, (4) function and data flow interaction, (5) final decision, and (6) security improvements. These steps were designed based on established security analysis methodologies and prior works on vulnerability detection frameworks, incorporating best practices from both program analysis and LLM reasoning strategies. By breaking down the reasoning process, this prompt aims to enhance model interpretability and logical consistency.

Our \textbf{Think Prompt} is inspired by Cognitive Science’s dual-process theory \cite{kahneman2011thinking, evans2008dual} and the Meta-CoT framework \cite{xiang2025towards}, both of which emphasize structured, deliberate reasoning akin to System 2 thinking. While traditional CoT facilitates step-by-step reasoning, it lacks the explicit organization and reflective verification necessary for complex problem-solving. Our Think Prompt addresses this by structuring responses into predefined sections such as \texttt{thinking} and \texttt{vulnerability\_assessment}, ensuring that LLMs systematically articulate their reasoning process before reaching a conclusion. This structured approach enhances interpretability, aligns with cognitive models of deliberate, strategic reasoning, and improves the model’s ability to analyze vulnerabilities with greater depth and consistency.

The \textbf{Think \& Verify Prompt} further extends structured reasoning by introducing a two-phase verification process. First, the model performs an initial vulnerability detection pass. Then, it reassesses its conclusions by verifying its reasoning, assigning a confidence score (0 - 100\%), and providing a severity rating of the detected issue. This additional verification step is designed to mitigate false positives and improve assessment reliability, ensuring that models apply a more rigorous validation process before finalizing vulnerability classifications \cite{xiang2025towards}.

CoT, Think, and Think \& Verify primarily differ in their depth of reasoning and verification mechanisms. CoT structures reasoning into sequential analytical steps but does not require explicit documentation of thought processes. Think enhances this by enforcing structured reasoning documentation, ensuring models explicitly record their logical process. Think-Verify takes this further by introducing an additional verification phase, reinforcing model confidence. These distinctions allow us to evaluate the impact of structured reasoning, explicit self-assessment, and confidence-based validation on LLM performance in vulnerability detection. The detailed prompt templates are provided in the \hyperref[appendix]{Appendix section}.

\subsection{Evaluation Process}
The VulnSage evaluation framework leverages DSPy \cite{khattab2023dspy} to systematically optimize LLM prompts for vulnerability detection. This framework guides a reasoning LLM to interpret analyses and generate consistent labels across evaluation methods. The evaluation is conducted on two distinct tasks: vulnerability detection, where the model determines if a given code snippet contains a security flaw, and patch verification, where it assesses whether a code modification constitutes a security fix. This distinction enables a thorough analysis of LLMs' ability to differentiate between vulnerable code and its corresponding patches, providing a more comprehensive understanding of reasoning and decision-making capabilities.

Using accuracy as our primary metric, we assess both tasks by applying standardized decision labels for CoT, Think, and Think-Verify approaches: 1 (vulnerable), 0 (non-vulnerable), and 2 (not sure). The ambiguous label captures cases where the LLM either explicitly indicates uncertainty, requires additional context or information, or fails to adhere precisely to our prompt instructions. To enhance reproducibility, we set the temperature to 0.7—lower than default settings—balancing reduced randomness with preserved reasoning diversity. This approach addresses LLMs' inherent non-determinism, as even zero-temperature settings cannot guarantee consistent outputs in code analysis tasks \cite{ouyang2025empirical}.

\begin{table*}[h!]
  \centering
  \renewcommand{\arraystretch}{0.7}
  \caption{Model Performance Accuracy Across Prompt Strategies. Bold values show the best performance in each column. \underline{Underlined} values highlight the second-best performance in each column for each type of model.}
  \scriptsize
  \label{tab:overalacc}
  \resizebox{\textwidth}{!}{
  \begin{tabular}{ll|cc|cc|cc|cc|cc}
  \toprule
  \multicolumn{2}{c|}{Prompt} & \multicolumn{2}{c|}{Baseline} & \multicolumn{2}{c|}{CoT} & \multicolumn{2}{c|}{Think} & \multicolumn{2}{c|}{Think \& Verify} & \multicolumn{2}{c}{Average} \\
  \toprule
  Model & Context Length & Vuln & Patch & Vuln & Patch & Vuln & Patch & Vuln & Patch & Vuln & Patch \\
  \midrule
  \rowcolor{blue!15} \multicolumn{12}{c}{General Models} \\ 
  Deepseek-v2 (16B) & \textit{131K} & \textbf{93.93} & 2.87 & \underline{30.78} & \underline{68.63} & 38.03 & \textbf{67.54} & 56.07 & \textbf{65.35} & 54.70 & 51.10 \\
  Llama3.1 (8B) & \textit{131K} & \underline{20.57} & \underline{40.47} & 25.04 & \textbf{74.87} & \underline{53.79} & 43.51 & \underline{58.68} & \underline{39.29} & 39.52 & 49.54 \\
  Gemma2 (9B) & \textit{8K} & 5.23 & \textbf{46.88} & \textbf{64.67} & 34.91 & \textbf{68.97} & \underline{45.45} & \textbf{63.49} & 36.51 & 50.59 & 40.94 \\
  \rowcolor{gray!15} Average & & 39.91 & 30.07 & 40.16 & 59.47 & 53.60 & 52.17 & 59.41 & 47.05 & 48.27 & 47.19 \\
  \midrule
  \rowcolor{blue!15} \multicolumn{12}{c}{Code-Specific Models} \\ 
  Deepseek-coder-v2 (16B) & \textit{163K} & \textbf{88.36} & 6.91 & 51.85 & 49.16 & 53.88 & 47.98 & 54.22 & \underline{65.51} & \underline{62.08} & 42.39 \\
  Qwen2.5-coder (7B) & \textit{32K} & 0.34 & \textbf{49.92} & 44.86 & \underline{57.00} & 43.84 & \underline{57.76} & 55.31 & 52.28 & 36.09 & 54.24 \\
  Qwen2.5-coder (32B) & \textit{32K} & 1.01 & \underline{49.75} & 32.21 & \textbf{72.01} & 38.79 & \textbf{70.24} & 53.04 & \textbf{68.55} & 31.26 & \textbf{65.14} \\
  Codellama (7B) & \textit{16K} & \underline{22.34} & 39.04 & \textbf{80.27} & 25.21 & \textbf{81.37} & 28.33 & 70.83 & 52.70 & \textbf{63.70} & 36.32 \\
  Codellama (34B) & \textit{16K} & 3.29 & 47.47 & \underline{52.78} & 44.77 & \underline{72.93} & 38.62 & \textbf{72.26} & 50.59 & 50.32 & 45.36 \\
  \rowcolor{gray!15} Average & & 23.07 & 38.62 & 52.39 & 49.63 & 58.16 & 48.59 & 61.13 & 57.93 & 48.69 & 48.69 \\
  \midrule
  \rowcolor{blue!15} \multicolumn{12}{c}{Reasoning Models} \\ 
  Deepseek-R1 (7B) & \textit{131K} & \textbf{67.12} & 19.06 & \textbf{41.82} & 59.02 & \textbf{67.45} & 44.86 & \textbf{52.36} & 72.85 & 57.19 & 48.95 \\
  Deepseek-R1 (32B) & \textit{131K} & 64.76 & \textbf{38.45} & 25.04 & \textbf{73.02} & 33.98 & \textbf{70.07} & 45.03 & \textbf{73.36} & 42.20 & \underline{63.73} \\
  \rowcolor{gray!15} Average & & 65.94 & 28.76 & 33.43 & 66.02 & 50.72 & 57.47 & 48.70 & 73.11 & 49.70 & 56.34 \\
  \midrule
  \rowcolor{gray!15} Overall Average & & 36.70 & 33.98 & 44.93 & 54.86 & 55.46 & 51.64 & 57.94 & 57.70 & 48.76 & 49.54 \\
  \bottomrule
  \end{tabular}
  } 
\end{table*}

\section{Results and Discussion}
\label{section:result}

We evaluate ten LLMs across three categories based on their specialized capabilities: general-purpose models (Deepseek-v2, Llama3.1, Gemma2), code-specialized models (Deepseek-coder, Qwen-coder, Codellama), and reasoning-focused models (Deepseek R1). General models serve as baselines, code-specific models are optimized for programming tasks, and reasoning models excel in logical analysis and step-by-step problem-solving. Using four prompting strategies—Baseline, CoT, Think, and Think \& Verify—we assess their performance in vulnerability detection and patch verification, gaining insights into how structured reasoning affects accuracy.

Our analysis consists of four key evaluations: (1) assessing model accuracy for vulnerability detection and patch verification across different prompts and comparing performance with a leakage-free dataset to determine whether LLMs rely on genuine reasoning or prior exposure; (2) evaluating robustness against dataset noise by introducing irrelevant modifications and measuring performance degradation across different prompting strategies; (3) investigating biases across CWE types, analyzing variations in detection accuracy for different vulnerability classes; and (4) exploring the impact of code granularity by examining how detection performance varies when models process isolated functions, entire files, or multi-file interactions. This comprehensive evaluation provides deeper insights into the strengths and weaknesses of LLMs in real-world SVD.

\subsection{Overall LLM Performance}

We evaluated the overall effectiveness of different LLMs in detecting software vulnerabilities and verifying patches across the various prompting strategies. Table~\ref{tab:overalacc} presents the accuracy for each model-prompt combination, where each number represents the percentage of samples correctly identified by the LLM. Specifically, the "Vuln" column shows accuracy in correctly determining if a code snippet contains a vulnerability, and the "Patch" column indicates accuracy in verifying whether a code modification effectively resolves a vulnerability.
The results highlight several key insights:

\textbf{Prompt Strategy.} The Think \& Verify strategy consistently outperforms the Baseline. For instance, vulnerability detection accuracy improves from an average of 36.7\% (Baseline) to 57.94\% (Think \& Verify), reflecting a 21.24\% point improvement. Similarly, patch verification accuracy rises from 33.98\% to 57.70\%, a 23.72 percentage point increase. These gains suggest structured reasoning combined with explicit verification substantially enhances LLMs' ability to identify vulnerabilities and confirm patches. Notably, reasoning models achieve the highest accuracy in patch verification, with Deepseek R1 32B reaching 73.36\%, although these models perform less effectively in vulnerability detection tasks. Conversely, Codellama 34B excels in vulnerability detection under the Think \& Verify strategy, achieving an accuracy of 72.26\%. 

This contrast highlights the differing strengths of reasoning-based and code-specialized models. Patch verification benefits from structured logical reasoning to assess whether a fix adequately addresses a vulnerability, aligning well with models like Deepseek R1. In contrast, vulnerability detection relies more on pattern recognition and deep code understanding, where code-specialized models such as Codellama, trained extensively on programming data, excel. The Think \& Verify strategy further enhances both tasks by encouraging self-assessment, reducing false positives in patch verification and improving model confidence in vulnerability identification.
    
\textbf{Model Performance.} Code-specialized models consistently achieve higher accuracy in both vulnerability detection and patch verification tasks compared to general-purpose and reasoning-focused models. Notably, Codellama 7B attains the highest average vulnerability detection accuracy across all four prompting strategies at 63.70\%, while Qwen2.5-coder 32B leads in patch verification accuracy with an average of 65.14\%. These results indicate that models specifically trained on programming data demonstrate superior capability in accurately identifying vulnerabilities and verifying patches, outperforming general-purpose models like Gemma2 9B (50.59\%) and Llama3.1 8B (39.52\%) as well as reasoning-oriented models such as Deepseek R1 7B (57.19\%). 

Distinct patterns emerged when analyzing specific vulnerabilities. For instance, when evaluating a buffer overflow vulnerability (CWE-119, CVE-2016-1953), only Codellama-7B and Deepseek-coder-v2-16B successfully identified it, effectively leveraging structured reasoning via the Think \& Verify prompt despite not being primarily reasoning-oriented architectures. An illustrative code example highlighting this finding is provided in Listing \ref{fig:code1}. Further details on performance across specific CWE categories are presented in Section \ref{cwe_distribution}
    
\textbf{Context Length Size.} The impact of context length size on model performance reveals complex patterns rather than a simple linear relationship. Models with larger context windows (131K-163K tokens) demonstrate superior performance in several scenarios, particularly with structured reasoning prompts. For instance, Deepseek-R1 models (131K context) achieve the highest patch verification accuracy with Think \& Verify prompts (72.85\% and 73.36\%), suggesting that larger context windows enable more thorough analysis when verifying security fixes across multiple code components. However, context size alone does not determine performance—Gemma2 (9B) with only an 8K window achieves impressive vulnerability detection results with Think (68.97\%) and Think \& Verify (63.49\%), outperforming some larger-context counterparts. Similarly, Codellama (7B) with a 16K context window attains 81.37\% vulnerability detection accuracy with Think prompting, surpassing many larger-context models. These findings suggest that while larger context windows generally benefit complex reasoning tasks like patch verification, model architecture and training can sometimes compensate for context limitations in certain vulnerability detection scenarios.
    
\textbf{Ambiguous Cases:} As shown in Table \ref{tab:ambiguous_responses}, models frequently express uncertainty when analyzing complex vulnerabilities, indicated by "not sure" responses (value 2). The frequency of these ambiguous cases varies notably by prompting strategy. On average, CoT prompt yielded approximately 120.3 ambiguous results for vulnerability detection and 120.9 for patch verification (20.3\% of all cases), while Think prompting reduced uncertainty for vulnerability detection to 95.5 cases but maintained higher ambiguity for patch verification (109.2). In contrast, Think-Verify significantly reduced uncertainty to just 50.6 ambiguous vulnerability detection cases and 57.3 patch verification cases (9.1\% overall), demonstrating how structured verification helps models reach better conclusions. These ambiguous responses typically occur when analyzing vulnerabilities involving complex inter-component interactions or subtle security patterns that challenge even specialized models, but the additional verification step in Think-Verify substantially improves decision confidence.

\begin{table}[htbp]
    \centering
    \caption{Number of Ambiguous Results}
    \renewcommand{\arraystretch}{0.7}
    \resizebox{\columnwidth}{!}{%
    \begin{tabular}{l|cc|cc|cc}
        \hline
        \multirow{2}{*}{\textbf{Model}} & \multicolumn{2}{c|}{\textbf{CoT}} & \multicolumn{2}{c|}{\textbf{Think}} & \multicolumn{2}{c}{\textbf{Think \& Verify}} \\
        \cline{2-7}
        & Vuln & Patch & Vuln & Patch & Vuln & Patch \\
        \hline
        \rowcolor{blue!15} \multicolumn{7}{c}{General Models} \\ 
        Deepseek-v2 (16B) & 161 & 160 & 87 & 117 & 59 & 63 \\
        Llama3.1 (8B) & 39 & 34 & 18 & 12 & 66 & 66 \\
        Gemma2 (9B) & 139 & 138 & 190 & 239 & 1 & 1 \\
        \rowcolor{gray!15} Average & 113.0 & 110.7 & 98.3 & 122.7 & 42.0 & 43.3 \\
        \hline
        \rowcolor{blue!15} \multicolumn{7}{c}{Code-Specific Models} \\ 
        Deepseek-coder (16B) & 135 & 143 & 45 & 51 & 51 & 51 \\
        Qwen2.5-coder (7B) & 146 & 90 & 38 & 49 & 18 & 14 \\
        Qwen2.5-coder (32B) & 110 & 110 & 48 & 53 & 49 & 51 \\
        Codellama (7B) & 62 & 73 & 81 & 122 & 52 & 61 \\
        Codellama (34B) & 106 & 145 & 129 & 138 & 63 & 62 \\
        \rowcolor{gray!15} Average & 111.8 & 112.2 & 68.2 & 82.6 & 46.6 & 47.8 \\
        \hline
        \rowcolor{blue!15} \multicolumn{7}{c}{Reasoning Models} \\ 
        Deepseek-R1 (7B) & 158 & 162 & 148 & 162 & 71 & 70 \\
        Deepseek-R1 (32B) & 147 & 154 & 171 & 149 & 76 & 134 \\
        \rowcolor{gray!15} Average & 152.5 & 158.0 & 159.5 & 155.5 & 73.5 & 102.0 \\
        \hline
        \rowcolor{gray!15} \textbf{Overall Average} & \textbf{120.3} & \textbf{120.9} & \textbf{95.5} & \textbf{109.2} & \textbf{50.6} & \textbf{57.3} \\
        \hline
    \end{tabular}%
    }
    \label{tab:ambiguous_responses}
\end{table}

\textbf{Data Leakage.} To assess potential training data leakage, we created two distinct datasets. The primary dataset (Table \ref{tab:overalacc}) includes vulnerabilities spanning from 2002 to 2019, which could potentially overlap with the training data of evaluated LLMs. In contrast, the second dataset (Table \ref{tab:dataleakage}) exclusively comprises vulnerabilities discovered after the training cutoff dates of all evaluated models, thereby ensuring a genuine zero-shot evaluation scenario. The comparison between these datasets reveals that LLMs rely primarily on reasoning rather than memorization when detecting vulnerabilities. For instance, the Think \& Verify prompt strategy achieved higher vulnerability detection accuracy (67.5\%) on the leakage-free dataset compared to the potentially contaminated dataset (57.94\%). Similarly, structured reasoning prompts such as CoT improved accuracy by approximately 7.07 percentage points, indicating their effectiveness even on previously unseen vulnerabilities.

These findings clearly illustrate that structural and functional changes introduced in newer codebase versions were not memorized by the models. Instead, the models demonstrated effective reasoning capabilities, validating their generalization in detecting and verifying security vulnerabilities in evolving real-world software.

\begin{table}[t]
  \centering
  \renewcommand{\arraystretch}{0.7}
  \caption{Accuracy of LLMs on Vulnerabilities Discovered in 2025 (Post-Training Cut-off)}
  \label{tab:dataleakage}
  \resizebox{\columnwidth}{!}{
  \begin{tabular}{l|cc|cc|cc|cc}
  \toprule
  \multicolumn{1}{c|}{Prompt} & \multicolumn{2}{c|}{\textbf{Baseline}} & \multicolumn{2}{c|}{\textbf{CoT}} & \multicolumn{2}{c|}{\textbf{Think}} & \multicolumn{2}{c}{\textbf{Think \& Verify}} \\
  \toprule
  Model & Vuln & Patch & Vuln & Patch & Vuln & Patch & Vuln & Patch \\
  \midrule
  \rowcolor{blue!15} \multicolumn{9}{c}{General Models} \\ 
  Deepseek-v2 (16B) & 90 & 10 & 10 & 70 & 40 & 100 & 80 & 85 \\
  Llama3.1 (8B) & 20 & 80 & 20 & 80 & 70 & 40 & 70 & 40 \\
  Gemma2 (9B) & 10 & 60 & 70 & 25 & 70 & 45 & 80 & 45 \\
  \rowcolor{gray!15} Average & 40.0 & 50.0 & 33.3 & 58.3 & 60.0 & 61.7 & 76.7 & 56.7 \\
  \midrule
  \rowcolor{blue!15} \multicolumn{9}{c}{Code-Specific Models} \\ 
  Deepseek-coder-v2 (16B) & 60 & 20 & 70 & 40 & 65 & 20 & 60 & 20 \\
  Qwen2.5-coder (7B) & 0 & 80 & 45 & 70 & 60 & 20 & 60 & 80 \\
  Qwen2.5-coder (32B) & 10 & 80 & 75 & 40 & 70 & 45 & 55 & 70 \\
  Codellama (7B) & 30 & 70 & 65 & 30 & 75 & 20 & 75 & 30 \\
  Codellama (34B) & 0 & 100 & 65 & 55 & 100 & 30 & 80 & 10 \\
  \rowcolor{gray!15} Average & 20.0 & 70.0 & 64.0 & 47.0 & 74.0 & 27.0 & 66.0 & 42.0 \\
  \midrule
  \rowcolor{blue!15} \multicolumn{9}{c}{Reasoning Models} \\ 
  Deepseek-R1 (7B) & 50 & 80 & 50 & 50 & 75 & 65 & 55 & 60 \\
  Deepseek-R1 (32B) & 70 & 30 & 50 & 70 & 40 & 80 & 60 & 70 \\
  \rowcolor{gray!15} Average & 60.0 & 55.0 & 50.0 & 60.0 & 57.5 & 72.5 & 57.5 & 65.0 \\
  \midrule
  \rowcolor{gray!15} Overall Average & \textbf{34.0} & \textbf{61.0} & \textbf{52.0} & \textbf{53.0} & \textbf{67.5} & \textbf{46.5} & \textbf{67.5} & \textbf{51.0} \\
  \bottomrule
  \end{tabular}
  }
\end{table}

\begin{figure}[ht]
\centering
\begin{lstlisting}[style=cstylediff,
                  caption={CVE-2016-1953 Fix: Adding null checks before adding \texttt{mObserver} to prevent null pointer dereference in Mozilla's Windows gamepad code.}, 
                  label={fig:code1}]
// File: dom/gamepad/windows/WindowsGamepad.cpp
void WindowsGamepadService::Cleanup() {
  if (mXInputPollTimer) {
    mXInputPollTimer->Cancel();
  }
  mGamepads.Clear();
(*@\setlength{\fboxsep}{0pt}\colorbox{addbg}{\strut\hspace*{-\fboxsep}\,\,if (mObserver) \{\hfill\hspace*{-\fboxsep}}@*)
(*@\setlength{\fboxsep}{0pt}\colorbox{addbg}{\strut\hspace*{-\fboxsep}\,\,\,\,mObserver->Stop();\hfill\hspace*{-\fboxsep}}@*)
(*@\setlength{\fboxsep}{0pt}\colorbox{addbg}{\strut\hspace*{-\fboxsep}\,\,\,\,mObserver = nullptr;\hfill\hspace*{-\fboxsep}}@*)
(*@\setlength{\fboxsep}{0pt}\colorbox{addbg}{\strut\hspace*{-\fboxsep}\,\,\}\hfill\hspace*{-\fboxsep}}@*)
}

void WindowsGamepadService::DevicesChanged(DeviceChangeType type) {
  if (type == DeviceChangeNotification) {
(*@\setlength{\fboxsep}{0pt}\colorbox{delbg}{\strut\hspace*{-\fboxsep}\,\,\,\,mObserver->SetDeviceChangeTimer();\hfill\hspace*{-\fboxsep}}@*)
(*@\setlength{\fboxsep}{0pt}\colorbox{addbg}{\strut\hspace*{-\fboxsep}\,\,\,\,if (mObserver) \{\hfill\hspace*{-\fboxsep}}@*)
(*@\setlength{\fboxsep}{0pt}\colorbox{addbg}{\strut\hspace*{-\fboxsep}\,\,\,\,\,\,mObserver->SetDeviceChangeTimer();\hfill\hspace*{-\fboxsep}}@*)
(*@\setlength{\fboxsep}{0pt}\colorbox{addbg}{\strut\hspace*{-\fboxsep}\,\,\,\,\}\hfill\hspace*{-\fboxsep}}@*)
  } else if (type == DeviceChangeStable) {
    ScanForDevices();
  }
}
\end{lstlisting}
\end{figure}


\subsection{Impact of Data Noise}
\label{sec:impactofnoise}
\begin{figure*}[htbp]
    \centering
    \includegraphics[width=1\textwidth]{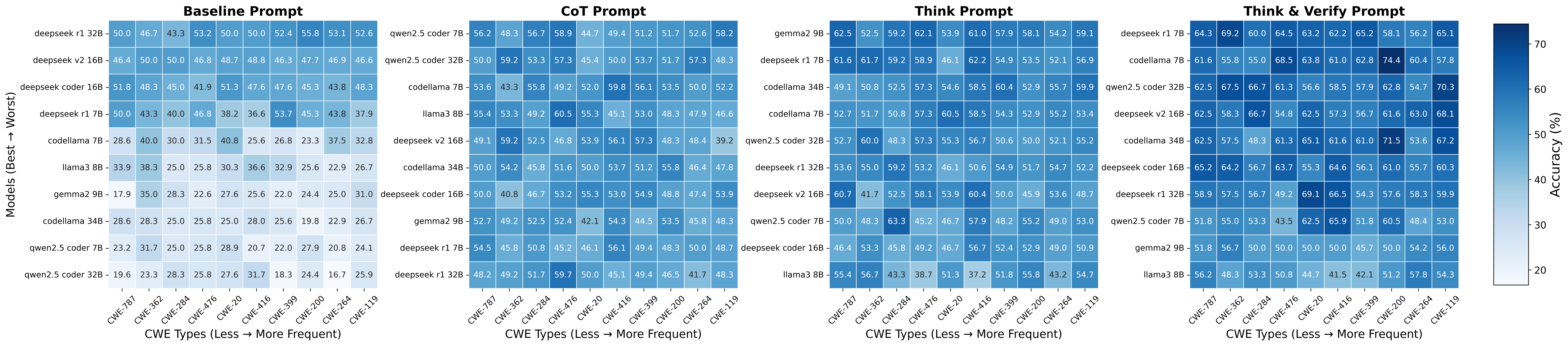}
    \tiny
    \caption{Aggregated accuracy of vulnerability detection and patch verification for the top 10 most frequent CWEs in our dataset, which align with commonly occurring vulnerabilities in real-world C and C++ system projects. Each heatmap represents a different prompting strategy, showing how accuracy varies across CWE types.} 
    \label{fig:top_cwe}
\end{figure*}

\begin{figure}[htbp]
    \centering
    \includegraphics[width=1.0\columnwidth]{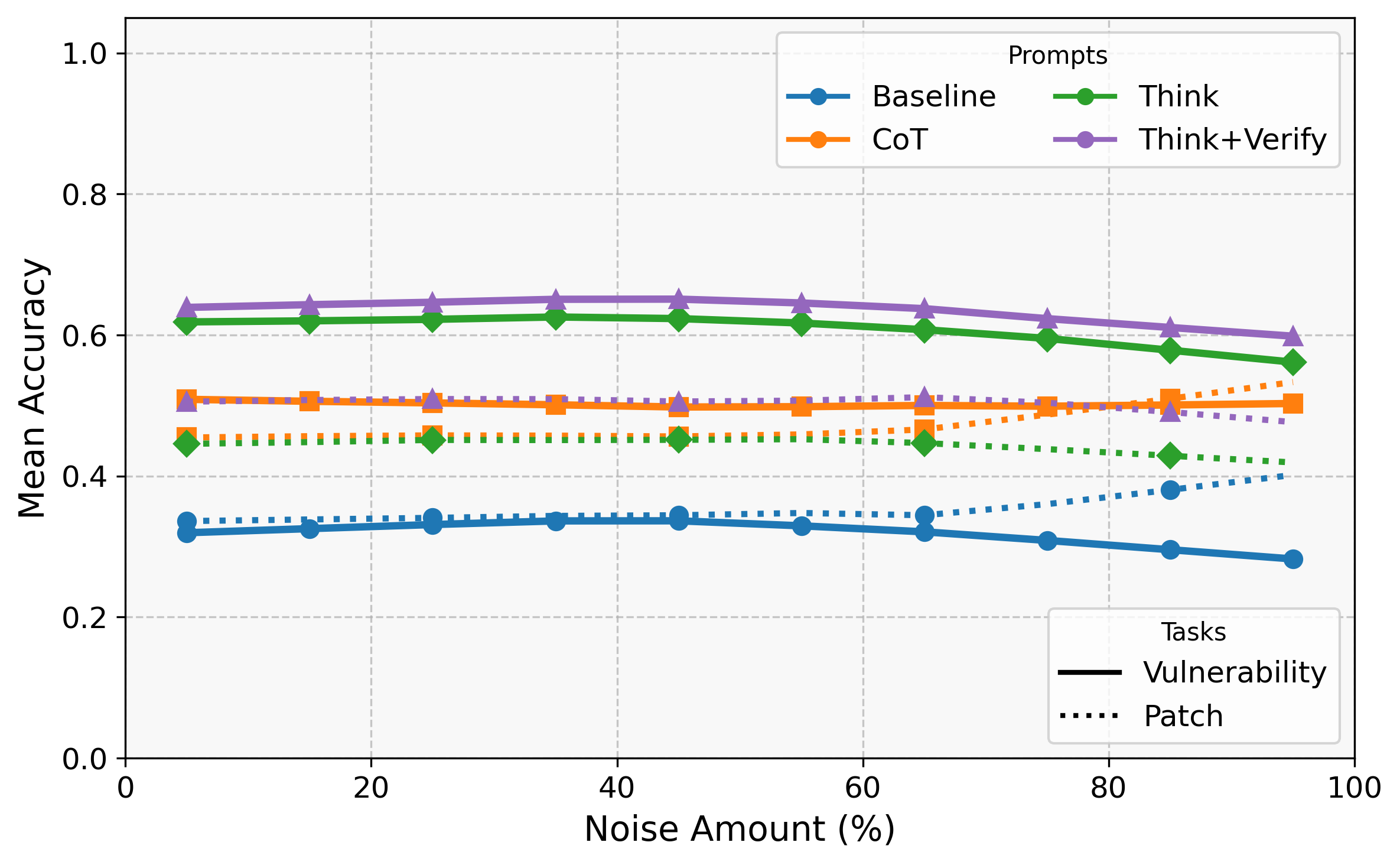}
    \tiny
    \caption{\textbf{Correlation between Noise and LLMs Performance.} The figure shows how different prompting strategies perform on vulnerability detection (solid lines) and patch verification (dotted lines) tasks as noise in the input code increases from 0\% to 100\%, with each line representing the mean accuracy across all evaluated models.}
    \label{fig:noise}
\end{figure}

Noise in training and testing data is an inherent challenge in SVD, often leading to misleading or ambiguous signals that can degrade model performance. However, as illustrated in Figure~\ref{fig:noise}, the impact of noise on LLM accuracy remains relatively modest across all prompting strategies. While an overall decline in accuracy is observed as noise levels increase, the rate of deterioration varies significantly across different strategies.

Models employing the Think \& Verify and Think strategies demonstrate superior robustness, maintaining a relatively stable accuracy despite increasing noise.  At the highest noise levels, the accuracy of these strategies declines by only 2.2\% and 2.9\%, respectively, indicating their ability to mitigate the effects of irrelevant modifications and extract essential security-related patterns. In contrast, Baseline strategies exhibit a more pronounced sensitivity to noise, with accuracy decreasing by approximately 6.4\%, suggesting that less structured reasoning approaches struggle to differentiate meaningful security indicators from noise. Notably, the CoT strategy remains largely unaffected by noise, suggesting that its step-by-step analysis provides stability despite the lack of explicit verification.

Interestingly, a divergent trend is observed between vulnerability detection and patch verification tasks. While vulnerability detection accuracy consistently declines as noise increases, patch verification accuracy exhibits a slight upward trend, particularly in the Baseline and CoT strategies. This pattern suggests that excessive modifications—such as added or deleted lines—may introduce contextual cues that assist in distinguishing patched code from vulnerable code, thereby improving patch verification accuracy.

\subsection{CWE Distributions}
\label{cwe_distribution}
The analysis of model performance across different CWE types (Figure \ref{fig:top_cwe}) reveals key insights into LLMs' vulnerability detection capabilities. Our dataset includes critical real-world vulnerabilities, with the most frequent being memory safety issues like CWE-119 (Buffer Overflow, 58 instances) and CWE-416 (Use-After-Free, 41 instances), as well as access control vulnerabilities such as CWE-264 (Permissions, 48 instances) and CWE-284 (Access Control Violation, 30 instances).

Each prompting strategy shows distinct performance patterns across CWE categories. The Baseline prompt generally yields the lowest performance, with Deepseek R1 32B showing consistently stronger results across multiple CWEs compared to other models, particularly for CWE-399 and CWE-200. In contrast, models like Qwen2.5 Coder 32B and Codellama 34B demonstrate notably weaker performance under this basic approach. The Chain-of-Thought approach substantially improves results for most models, with Qwen2.5 Coder 7B showing particularly strong performance on CWE-787 and CWE-476, while Codellama 7B excels in CWE-416.
\begin{figure*}[htbp]
    \centering
    \includegraphics[width=1\textwidth]{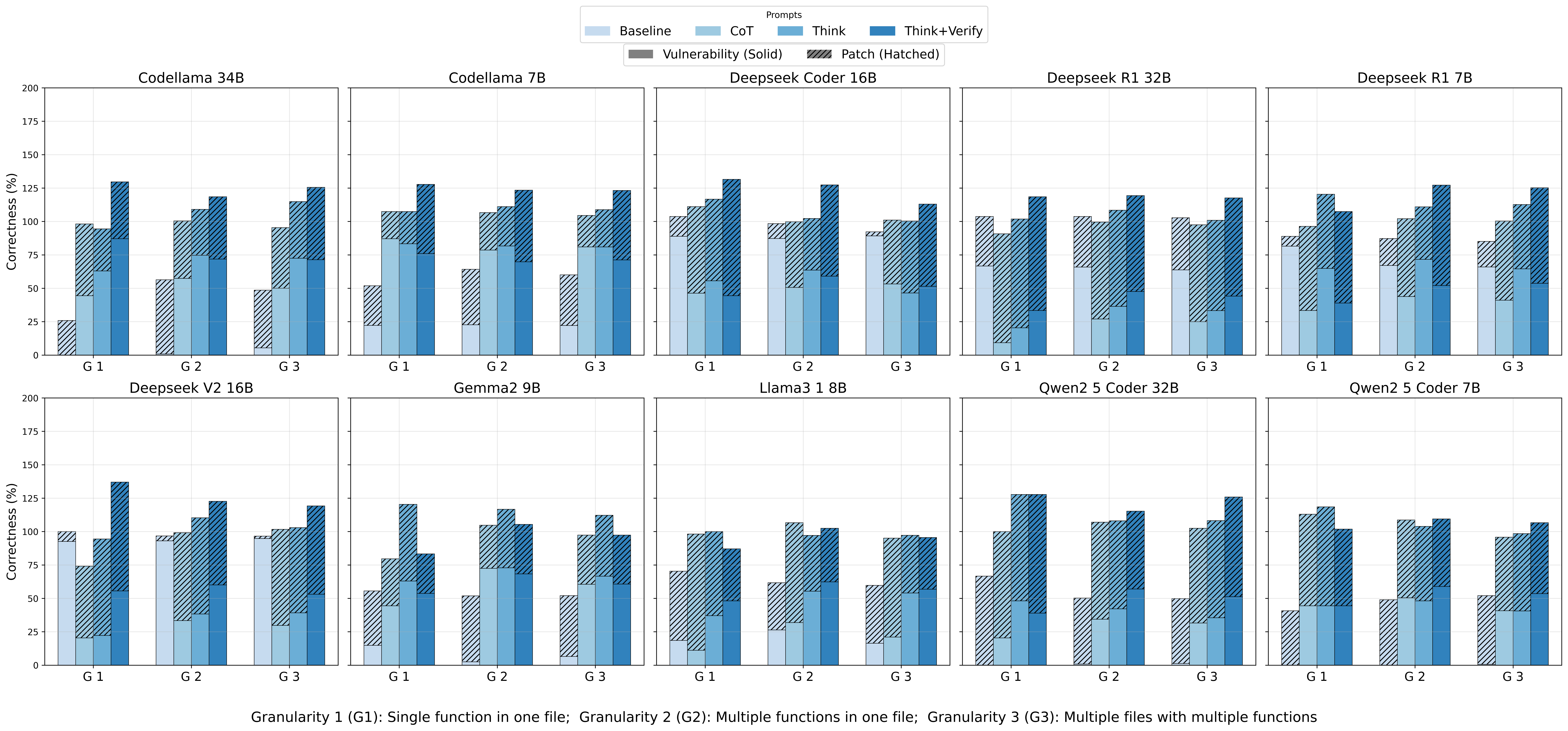}
    \caption{Impact of Granularity on LLM Performance Across Prompt Strategies. The figure shows vulnerability detection (solid bars) and patch verification (hatched bars) correctness. The y-axis represents aggregated accuracy, where stacked bars sum both tasks (e.g., 100\% in each totals 200\%).} 
    \label{fig:granularity}
\end{figure*}
The structured reasoning prompts yield further improvements. With the Think strategy, Gemma2 9B achieves impressive results across multiple vulnerability types, particularly for CWE-787 (62.5\%) and CWE-476 (62.1\%), while Llama3 8B lags behind at 48.8\%. This outcome indicates that some models benefit substantially from an organized internal reasoning process, but there is variability in how effectively each model exploits this prompt style. Deepseek R1 7B also performs strongly with this approach, especially on CWE-362 (61.7\%) and CWE-416 (62.2\%). Think \& Verify delivers the best overall results for most models, with Deepseek R1 7B showing exceptional performance across nearly all CWE types, reaching 69.2\% for race conditions (CWE-362) and 65.1\% for buffer overflows (CWE-119). Similarly, Qwen2.5 Coder 32B achieves an impressive 70.3\% on CWE-119 with this approach. The addition of a verification step appears especially beneficial for categories such as CWE‑119 and CWE‑362, where reasoning about memory or concurrency may require additional confirmation to avoid misclassification.

Looking at individual CWEs, we observe distinct performance patterns that reveal how different model architectures handle various vulnerability types. CWE-119 (Buffer Overflow) and CWE-399 (Resource Management Errors) often show substantial gains from more advanced prompting strategies (CoT, Think, and Think \& Verify), with code-specific models like Codellama and Deepseek-Coder excelling particularly on memory safety issues (CWE-119, CWE-787). In contrast, reasoning-focused models such as Deepseek R1 perform exceptionally well on concurrency vulnerabilities (CWE-362) when paired with structured verification approaches. Despite these targeted strengths, vulnerabilities such as CWE-476 (NULL Pointer Dereference) still pose challenges for many models across all prompting strategies.

Memory management vulnerabilities are often particularly challenging to detect, as illustrated in Listing \ref{fig:code2}, where a subtle change from raw to smart pointers addresses a critical Use-After-Free vulnerability. Only Codellama 7B and 34B could answer this case accurately in the Think \& Verify strategy, which demonstrates how specialized code models can leverage structured verification to reason about memory safety even when the vulnerability indicators are implicit. This finding highlights the importance of both domain-specific model training and appropriate prompting techniques when addressing complex memory management issues that require deep language-specific understanding.

The diversity of best-performing models under different prompts—Deepseek R1 32B (Baseline), Qwen2.5 Coder 7B (CoT), Gemma2 9B (Think), and Deepseek R1 7B (Think \& Verify)—underscores that no single approach or architecture consistently excels across all vulnerability types. Instead, there is a clear prompt–model interaction effect, where the best strategy depends on the underlying model and the specific CWE category. Moreover, these performance trends are influenced by sample distribution: CWE-119, with the highest number of samples, contributes significantly to overall accuracy patterns, whereas categories with fewer instances (e.g., CWE-787) exhibit greater performance variability.

 
 
 
 

\begin{figure}[ht]
\centering
\begin{lstlisting}[style=cstylediff,
  caption={CVE-2012-4213 (CWE-416: Use-After-Free) in Mozilla's \texttt{nsEditor::FindNode}. The vulnerable version (red) uses a raw pointer for \texttt{candidate}, leading to potential use-after-free issues. The patched version (green) replaces it with an \texttt{nsCOMPtr}, a smart pointer that ensures automatic reference counting, preventing premature deallocation.},
  label={fig:code2}]
 // File: editor/libeditor/base/nsEditor.cpp 
 nsIContent* nsEditor::FindNode(nsINode *aCurrentNode,
                 bool aGoForward,
                 bool aEditableNode,
                 bool bNoBlockCrossing) {
   if (IsEditorRoot(aCurrentNode)) { 
     return nullptr;
   }

(*@\setlength{\fboxsep}{0pt}\colorbox{delbg}{\strut\hspace*{-\fboxsep}\,\,nsIContent* candidate = \hfill\hspace*{-\fboxsep}}@*)
(*@\setlength{\fboxsep}{0pt}\colorbox{addbg}{\strut\hspace*{-\fboxsep}\,\,nsCOMPtr<nsIContent> candidate = \hfill\hspace*{-\fboxsep}}@*)

   FindNextLeafNode(aCurrentNode, aGoForward, bNoBlockCrossing);

   if (!candidate) {
     return nullptr;
   }

   if (!aEditableNode || IsEditable(candidate)) {
     return candidate;
   }

   return FindNode(candidate, aGoForward, aEditableNode, bNoBlockCrossing);
 }
\end{lstlisting}
\end{figure}

\subsection{Impact of Code Block Granularity Level}

The complexity of code structure poses a significant challenge in automated vulnerability detection and remediation. This section examines how varying levels of code granularity—ranging from isolated functions to multi-file systems—affect the performance of LLMs in SVD. Specifically, we analyze performance trends across three granularity levels: G1 (single function in one file), G2 (multiple functions in one file), and G3 (multiple functions across multiple files).

Our analysis reveals a nuanced relationship between code granularity and LLM performance. In many cases, increasing granularity enhances vulnerability detection accuracy, but this effect is model-dependent. For instance, CodeLlama 34B with the "Think" prompt improves from 62.96\% (G1) to 74.80\% (G2) before slightly decreasing to 72.36\% (G3). This suggests that additional contextual information can enhance vulnerability detection capabilities by providing more execution flow details. Similarly, Deepseek V2 16B under baseline prompting improves from 92.59\% (G1) to 94.72\% (G3), indicating that some models remain highly effective even with complex, multi-component code.

This contradicts the intuitive assumption that increased code complexity universally hampers detection performance. Instead, results indicate that certain models leverage cross-component interactions effectively, improving their detection of security flaws in larger code segments. Listing \ref{fig:code3} illustrates this challenge with a use-after-free vulnerability (CVE-2017-6874) spanning multiple Linux kernel files, where atomic operations were improperly used for reference counting. Detecting such vulnerabilities requires models to understand how the atomic counter in the header file relates to its usage in concurrent operations—context that would be lost if analyzing either file in isolation. When examining this particular example, models using Think \& Verify prompting were better able to trace the connection between the atomic counter definition and its unsafe usage pattern. Larger code blocks offer crucial contextual clues for detecting such cross-component vulnerabilities but may overwhelm models that lack strategies to structure their reasoning about complex interactions.

\textbf{Granularity-Prompt Interaction Effect.} The impact of granularity is strongly influenced by prompting strategies. While baseline prompts exhibit inconsistent performance patterns across granularity levels, structured reasoning prompts such as Think and Think+Verify demonstrate more predictable trends. In fact, CodeLlama 7B with the "Think" prompt remains stable across different granularities (83.33\% at G1, 81.35\% at G2, and 81.21\% at G3), suggesting resilience to increased complexity. On the other hand, "Think \& Verify" shows mixed trends, with some models improving at higher granularities while others experience a decline, implying that explicit verification may not always be beneficial in highly complex contexts.

These results suggest that advanced prompting techniques influence how models process increasing complexity—some enable better reasoning across functions, while others assist in breaking down intricate code dependencies.

\begin{figure}[ht]
\centering
\begin{lstlisting}[style=cstylediff,
  caption={CVE-2017-6874 in Linux kernel: The patch replaces an atomic reference counter with a lock-protected integer to prevent race conditions leading to a use-after-free vulnerability. By ensuring that \texttt{ucounts->count} is properly decremented and checked under \texttt{ucounts\_lock}, the fix eliminates unsafe concurrent access.},
  label={fig:code3}]
 // File path: include/linux/user_namespace.h
(*@\setlength{\fboxsep}{0pt}\colorbox{delbg}{\strut\hspace*{-\fboxsep}\,\,atomic\_t count;\hfill\hspace*{-\fboxsep}}@*)
(*@\setlength{\fboxsep}{0pt}\colorbox{addbg}{\strut\hspace*{-\fboxsep}\,\,int count;\hfill\hspace*{-\fboxsep}}@*)

 // File: kernel/ucount.c
 static struct ucounts *get_ucounts(struct user_namespace *ns, kuid_t uid) {
   // ...initialization code...
   new->ns = ns; new->uid = uid;
(*@\setlength{\fboxsep}{0pt}\colorbox{delbg}{\strut\hspace*{-\fboxsep}\,\,atomic\_set(\&new->count, 0);\hfill\hspace*{-\fboxsep}}@*)
(*@\setlength{\fboxsep}{0pt}\colorbox{addbg}{\strut\hspace*{-\fboxsep}\,\,new->count = 0;\hfill\hspace*{-\fboxsep}}@*)
   // ...more code...
 }

 static void put_ucounts(struct ucounts *ucounts) {
   unsigned long flags;
(*@\setlength{\fboxsep}{0pt}\colorbox{delbg}{\strut\hspace*{-\fboxsep}\,\,if (atomic\_dec\_and\_test(\&ucounts->count)) \{\hfill\hspace*{-\fboxsep}}@*)
(*@\setlength{\fboxsep}{0pt}\colorbox{delbg}{\strut\hspace*{-\fboxsep}\,\,\,\,spin\_lock\_irqsave(\&ucounts\_lock, flags);\hfill\hspace*{-\fboxsep}}@*)
(*@\setlength{\fboxsep}{0pt}\colorbox{addbg}{\strut\hspace*{-\fboxsep}\,\,spin\_lock\_irqsave(\&ucounts\_lock, flags);\hfill\hspace*{-\fboxsep}}@*)
(*@\setlength{\fboxsep}{0pt}\colorbox{addbg}{\strut\hspace*{-\fboxsep}\,\,ucounts->count -= 1;\hfill\hspace*{-\fboxsep}}@*)
(*@\setlength{\fboxsep}{0pt}\colorbox{addbg}{\strut\hspace*{-\fboxsep}\,\,if (!ucounts->count)\hfill\hspace*{-\fboxsep}}@*)
     hlist_del_init(&ucounts->node);
(*@\setlength{\fboxsep}{0pt}\colorbox{delbg}{\strut\hspace*{-\fboxsep}\,\,\,\,spin\_unlock\_irqrestore(\&ucounts\_lock, flags);\hfill\hspace*{-\fboxsep}}@*)
(*@\setlength{\fboxsep}{0pt}\colorbox{addbg}{\strut\hspace*{-\fboxsep}\,\,else ucounts = NULL;\hfill\hspace*{-\fboxsep}}@*)
(*@\setlength{\fboxsep}{0pt}\colorbox{addbg}{\strut\hspace*{-\fboxsep}\,\,spin\_unlock\_irqrestore(\&ucounts\_lock, flags);\hfill\hspace*{-\fboxsep}}@*)
(*@\setlength{\fboxsep}{0pt}\colorbox{delbg}{\strut\hspace*{-\fboxsep}\,\,\,\,kfree(ucounts);\hfill\hspace*{-\fboxsep}}@*)
(*@\setlength{\fboxsep}{0pt}\colorbox{delbg}{\strut\hspace*{-\fboxsep}\,\,\}\hfill\hspace*{-\fboxsep}}@*)
(*@\setlength{\fboxsep}{0pt}\colorbox{addbg}{\strut\hspace*{-\fboxsep}\,\,kfree(ucounts);\hfill\hspace*{-\fboxsep}}@*)
 }
\end{lstlisting}
\end{figure}

\textbf{Vulnerability Detection vs. Patch Verification.} A key observation is the divergent impact of granularity on vulnerability detection versus patch verification. While higher granularity often improves vulnerability detection, patch verification performance tends to decline. For example, Deepseek V2 16B’s patch accuracy under baseline prompting drops from 7.41\% (G1) to just 1.86\% (G3), despite an increase in vulnerability detection accuracy. This suggests that vulnerability detection benefits from additional execution context, as models can analyze interactions between functions to better recognize security flaws. Conversely, patch verification is more challenging at higher granularities because models must evaluate the correctness of fixes across multiple interdependent components rather than generating new fixes, ensuring that security vulnerabilities are properly mitigated without introducing new risks. Therefore, patch verification requires a different approach than vulnerability detection, with increased complexity posing a greater challenge for maintaining logical consistency.

\section{Threats to Validity}
\label{section:threats}
Our evaluation framework faces several threats to validity. First, the data collection process inherently introduces noise, a challenge common to existing vulnerability datasets. To mitigate this, we automatically pre-filtered approximately 39\% of our initial data—excluding commits with excessive file changes or non-code modifications—without relying on subjective assumptions. We subsequently employed an LLM-based noise quantification approach, which, while introducing some bias due to model assumptions, is transparently integrated into our evaluation to contextualize performance impacts.
Second, our noise analyzer may not fully capture the extent of noise in the data. Since it is based on model-driven reasoning rather than manual inspection of every commit, it may overlook or misclassify certain non-security-related modifications. While we validated the noise scoring mechanism by manually reviewing 2\% of the dataset, which aligned well with our expectations, there is still the possibility that some noisy or mislabeled commits remain.
Finally, we utilized pre-trained LLMs in a zero-shot setting, selecting test data from distinct Linux kernel codebase versions to minimize direct overlap with training datasets. Despite these measures, the uncertain scope of LLM training corpora leaves room for potential indirect data leakage. To further address this concern, we incorporated a subset of samples (ten samples) released after the LLMs' training cutoff dates (starting from 2025), thereby strengthening the external validity of our findings. However, the limited number of samples for assessing data leakage remains a constraint, potentially affecting the generalizability of our conclusions.

\section{Conclusion}
\label{section:conclusion}
Our study demonstrates that structured reasoning significantly enhances LLMs' ability to detect vulnerabilities and verify patches, with the Think \& Verify strategy reducing ambiguity and improving accuracy. Code-specialized models consistently outperform general-purpose alternatives, yet performance varies across vulnerability types, highlighting that no single approach excels universally. By introducing VulnSage, a rigorously curated dataset and evaluation framework, we provide a robust benchmark for assessing LLMs in real-world security scenarios. Our findings emphasize the need for continued advancements in reasoning-driven approaches to improve the reliability and generalization of LLMs in software vulnerability detection.

\bibliographystyle{ACM-Reference-Format}
\bibliography{ref}

\clearpage
\twocolumn
\flushbottom
\section{Appendix}
\label{appendix}

This appendix details the prompt strategies used to evaluate LLMs in SVD. Each prompt assesses different reasoning mechanisms, ranging from binary classification to structured, multi-step analysis. Additionally, the noise evaluation prompt helps quantify non-security-related modifications in vulnerability-fixing commits.

\subsection{Noise Evaluation Prompt}
Our noise evaluation prompt is designed to quantitatively assess the proportion of non-security-related changes in vulnerability fix commits. This structured approach helps identify and measure potential noise in the dataset, ensuring that our evaluation focuses on genuine security fixes rather than unrelated code changes.

\begin{tcolorbox}[colback=white, colframe=black, title=Noise Evaluation Prompt]
Task: You are a security analyst tasked with evaluating the ``noise'' in a commit that fixes a vulnerability.

``Noise'' is defined as the proportion of changes that are not directly related to the core vulnerability fix.

You are provided with the following information:

\textbf{Commit Description:} \\
\verb|{commit_desc}|

\textbf{Git Diff:} \\
\verb|{commit_diff}|

Please follow these steps:

\begin{enumerate}
    \item Review the commit description to understand the intent behind the changes.
    \item Analyze the git diff to identify what modifications were made.
    \item Based on your analysis, estimate the overall level of noise on a scale from 0 to 100, where 0 means nearly all changes are essential and 100 means most changes are unrelated.
    \item Provide a step-by-step reasoning of your analysis.
    \item On a new line, output your final result exactly in the following format:
\end{enumerate}

NOISE\_AMOUNT: X \\
REASONING: [Your detailed explanation]
\end{tcolorbox}

\subsection{Baseline Strategy}
The baseline strategy represents our control test, requiring models to make binary decisions about vulnerability presence without detailed explanations. This approach helps establish a fundamental performance benchmark and evaluate whether more complex prompting strategies provide meaningful improvements over simple classification.
\begin{tcolorbox}[colback=bluebox, colframe=blue!60!black, 
title=\textbf{Baseline Strategy (YES/NO + CWE Only)}, 
sharp corners=south, boxrule=1pt, width=\linewidth]
\textbf{Prompt:}  
You are a security expert specialized in identifying software vulnerabilities in C code.

Analyze the following code and determine whether it contains any vulnerabilities.

\begin{lstlisting}[language=C, 
                  backgroundcolor=\color{codebg},
                  basicstyle=\footnotesize\ttfamily,
                  breaklines=true,
                  frame=none]
[Code block goes here]
\end{lstlisting}

Provide your response in \textbf{exactly} the following format:
\begin{enumerate}
    \item \textbf{Vulnerability Present?} (YES or NO)
    \item \textbf{If YES, list the relevant CWE ID(s) only} (e.g., CWE-119, CWE-79).
\end{enumerate}

\textbf{Do not provide any explanation or additional details.}
\end{tcolorbox}

\subsection{CoT Strategy}
The Chain of Thought strategy implements a structured analytical framework that guides models through specific reasoning steps. This methodology helps evaluate whether breaking down the analysis process improves detection accuracy and provides more reliable vulnerability assessments.
\begin{tcolorbox}[colback=redbox, colframe=red!60!black, 
title=\textbf{CoT Strategy}, 
sharp corners=south, boxrule=1pt, width=\linewidth]
\textbf{Prompt:}  
You are a security expert specialized in vulnerability detection. Analyze the following C code using a structured approach.

\textbf{Step-by-step analysis:}
\begin{enumerate}
    \item \textbf{Code Structure Analysis:} Identify key components, data flow, and possible security risks.
    \item \textbf{Attack Surface \& Risk Assessment:} Identify unsafe functions and risky patterns.
    \item \textbf{Interaction \& Exploitability:} Examine function interactions and attack feasibility.
    \item \textbf{CWE Pattern Matching:} Classify vulnerabilities according to CWE.
    \item \textbf{Final Decision:} Justify whether the code is vulnerable or not.
    \item \textbf{Suggested Security Improvements:} Recommend fixes and mitigations.
\end{enumerate}

\begin{lstlisting}[language=C, 
                  backgroundcolor=\color{codebg},
                  basicstyle=\footnotesize\ttfamily,
                  breaklines=true,
                  frame=none]
[Code block goes here]
\end{lstlisting}
\end{tcolorbox}

\subsection{Think Strategy}
The Think strategy enforces explicit documentation of the reasoning process through structured sections. This approach allows us to evaluate both the final detection outcome and the quality of the underlying analysis, providing insights into the model's decision-making process.
\begin{tcolorbox}[colback=greenbox, colframe=green!60!black, 
title=\textbf{Think Strategy}, 
sharp corners=south, boxrule=1pt, width=\linewidth]
\textbf{Prompt:}  
You are a security expert analyzing C code for vulnerabilities. Use the following structured approach:

\textbf{Thinking Process}
\begin{lstlisting}[language=C, 
                  backgroundcolor=\color{codebg},
                  basicstyle=\footnotesize\ttfamily,
                  breaklines=true,
                  frame=none]
<thinking>
- Identify potential security issues.
- Consider different attack scenarios.
- Examine function interactions and data flows.
- Question assumptions about input validation.
- Verify findings and rule out false positives.
- Document confidence levels.
</thinking>
\end{lstlisting}

\textbf{Vulnerability Assessment}
\begin{lstlisting}[language=C, 
                  backgroundcolor=\color{codebg},
                  basicstyle=\footnotesize\ttfamily,
                  breaklines=true,
                  frame=none]
<vulnerability_assessment>
- Identified vulnerabilities
- Associated CWE(s)
- Severity ratings
- Relevant evidence from the code
</vulnerability_assessment>
\end{lstlisting}

\begin{lstlisting}[language=C, 
                  backgroundcolor=\color{codebg},
                  basicstyle=\footnotesize\ttfamily,
                  breaklines=true,
                  frame=none]
[Code block goes here]
\end{lstlisting}
\end{tcolorbox}

\subsection{Think \& Verify Strategy}

\begin{tcolorbox}[colback=graybox, colframe=gray!60!black, 
title=\textbf{Think \& Verify Strategy}, 
sharp corners=south, boxrule=1pt, width=\linewidth]
\textbf{Prompt:}  
You are a security expert conducting an in-depth vulnerability assessment. Follow these steps:

\textbf{1. Initial Analysis (Up to 3 Attempts)}
\begin{lstlisting}[language=C, 
                  backgroundcolor=\color{codebg},
                  basicstyle=\footnotesize\ttfamily,
                  breaklines=true,
                  frame=none]
<thinking>
- Examine the code structure.
- Identify security vulnerabilities.
- Consider attack vectors.
- Document any uncertainties.
</thinking>
\end{lstlisting}

\textbf{Findings}
\begin{lstlisting}[language=C, 
                  backgroundcolor=\color{codebg},
                  basicstyle=\footnotesize\ttfamily,
                  breaklines=true,
                  frame=none]
<findings>
- List identified vulnerabilities.
</findings>
\end{lstlisting}

\textbf{Confidence Assessment}
\begin{lstlisting}[language=C, 
                  backgroundcolor=\color{codebg},
                  basicstyle=\footnotesize\ttfamily,
                  breaklines=true,
                  frame=none]
<confidence>
- Assign a confidence score (0-100%).
- If confidence is >= 90%, proceed to verification.
- If confidence is < 90%, reanalyze before verification.
</confidence>
\end{lstlisting}

\textbf{2. Verification (Required for High-Confidence Finding)}
\begin{lstlisting}[language=C, 
                  backgroundcolor=\color{codebg},
                  basicstyle=\footnotesize\ttfamily,
                  breaklines=true,
                  frame=none]
<verification>
- Validate each vulnerability.
- Check for false positives.
- Confirm CWE classification accuracy.
- Consider edge cases.
</verification>
\end{lstlisting}

\textbf{3. Final Assessment}
\begin{lstlisting}[language=C, 
                  backgroundcolor=\color{codebg},
                  basicstyle=\footnotesize\ttfamily,
                  breaklines=true,
                  frame=none]
<assessment>
- List verified vulnerabilities.
- Map to CWE.
- Assign severity ratings.
- Recommend security fixes.
</assessment>
\end{lstlisting}

\begin{lstlisting}[language=C, 
                  backgroundcolor=\color{codebg},
                  basicstyle=\footnotesize\ttfamily,
                  breaklines=true,
                  frame=none]
[Code block goes here]
\end{lstlisting}
\end{tcolorbox}

The Think \& Verify strategy represents our most comprehensive approach, incorporating multiple analysis passes and explicit verification steps. This method tests whether additional verification and confidence scoring can reduce false positives and improve the reliability of vulnerability detection.

\end{document}